# Investigation of MoS$_2$ and Graphene Nanosheets by Magnetic Force Microscopy**

*Hai Li,[1] Xiaoying Qi,[2] Jumiati Wu,[1] Zhiyuan Zeng,[1] Jun Wei,[2] Hua Zhang[1]\**

[1]School of Materials Science and Engineering, Nanyang Technological University, 50 Nanyang Avenue, Singapore 639798, Singapore

[2]Singapore Institute of Manufacturing Technology, 71 Nanyang Drive, Singapore 638075, Singapore.

\* To whom correspondence should be addressed.

Phone: +65-6790-5175. Fax: +65-6790-9081

E-mail: hzhang@ntu.edu.sg, hzhang166@yahoo.com

Website: http://www.ntu.edu.sg/home/hzhang/

**ABSTRACT.** For the first time, the magnetic force microscopy (MFM) is used to characterize the mechanically-exfoliated single- and few-layer $MoS_2$ and graphene nanosheets. By analysis of the phase and amplitude shifts, the magnetic response of $MoS_2$ and graphene nanosheets exhibits the dependence on their layer number. However, the solution-processed single-layer $MoS_2$ nanosheet shows the reverse magnetic signal to the mechanically-exfoliated one, and the graphene oxide nanosheet has not shown any detectable magnetic signal. Importantly, graphene and $MoS_2$ flakes become nonmagnetic when they exceed a certain thickness.

Owing to their unusual electronic structures and exceptional physical properties, two-dimensional (2D) layered nanomaterials have received numerous attention in recent years.[1-12] Graphene, a single-layer 2D carbon material with zero band gap, is the most studied 2D nanomaterials and shows extensive applications due to its fascinating properties, such as high electron mobility, good thermal conductivity, excellent elasticity and mechanical stiffness.[4, 5, 7, 9] Recently, the layered transition metal dichalcogenides (TMDs) have also attracted considerable interest due to their unique physical properties, such as the ideal band gap and large in-plane electron mobility.[1-3, 5, 6, 8, 13-19] Among them, $MoS_2$, one kind of the most stable layered TMDs, shows great applications in transistors,[8, 17, 20] sensors,[3, 21, 22] memory devices,[23, 24] and hydrogen evolution.[25, 26]

Compared to the extensive experimental and theoretical studies on electrical, mechanical and optical properties of atomically thin graphene and $MoS_2$ nanosheets, only a few studies on their magnetic properties have been reported.[27-34] It is known that the bulk graphite is diamagnetic and single crystal $MoS_2$ is nonmagnetic.[27, 35, 36] However, the atomically thin 2D nanosheet usually gives novel physical properties compared to its bulk material due to the quantum and surface effects. Recently, the room-temperature ferromagnetism of reduced graphene oxide (rGO) has been reported through the measurement by magnetometer.[28, 29, 32] In addition, the room-temperature magnetic properties of multi-layered functionalized epitaxial graphene on SiC wafers have also been measured by magnetic force microscopy (MFM) and scanning tunneling microscopy (STM).[37, 38] The $MoS_2$ thin film with typical edge dimension of ~100 nm has exhibited weak magnetism, which is attributed to the existence of the zigzag edges in the ferromagnetic ground state.[27, 31] As the thickness and grain size decrease, the increase of magnetism has been observed in both graphene and $MoS_2$ thin films.[27, 28, 31] To date, the mechanical exfoliation is still the easiest and fastest way to produce high-quality, atomically thin nanosheets of single-crystal 2D layered nanomaterials, which are suitable for fundamental studies.[2, 3, 5, 8, 12, 13, 16, 17, 31] Although the theoretical calculation can predict the magnetic property of individual single- or few-layer graphene and MoS$_2$ nanosheets,[32-34] to the best of our knowledge, there is no direct experimental study of the magnetic response of individual, mechanically-exfoliated, pristine single- and few-layer graphene and MoS$_2$ nanosheets.

Magnetic force microscopy (MFM) is a powerful tool to detect magnetic interactions between the magnetized AFM tip and nanostructured sample.[39-46] Since MFM can provide nanometer resolution similar to AFM and has ability to detect nanoscopic magnetic domains, it is able to distinguish the magnetic and nonmagnetic responses in the micro- and nanoscale. Therefore, MFM is desirable to characterize the magnetic response of single- or few-layer 2D nanosheets, such as graphene and MoS$_2$.

Herein, MFM is used to characterize the magnetic responses of mechanically-exfoliated single- and few-layer graphene and MoS$_2$ nanosheets by analysis of the phase and amplitude shifts. Negative phase shift, which represents the attractive interaction between magnetic tip and sample,[40, 42, 43] was observed in combination with the positive amplitude shift in both single- and few-layer graphene and MoS$_2$ nanosheets. In addition, we found that the magnetic response of graphene and MoS$_2$ nanosheets depends on their layer number. However, graphene shows different thickness-dependent magnetic response compared to MoS$_2$. Moreover, graphene and MoS$_2$ flakes become nonmagnetic when they exceed a certain thickness.

**RESULTS AND DISCUSSION**

In our MFM study, a two-pass tapping/lift mode is used to measure the relatively weak but long-range magnetic interactions in order to minimize the influence of sample topography.[42, 45] Each line in the image is scanned twice during the operation of MFM. After a flexible cantilever equipped with a magnetized tip scans over the surface of a sample to obtain the topographic information, the tip is raised up to a certain height (so-called lift height) above the sample surface to measure the magnetic response by monitoring the cantilever's frequency or phase shift in the lift mode scan. Phase shift is used to analyze the magnetic response in the present study due to its higher sensitivity compared to the frequency shift (Figure S1 in Supporting Information (SI)).

Figure 1A shows the optical image of a mechanically-exfoliated $MoS_2$ flake on a 90 nm $SiO_2$ coated Si substrate, referred to 90 nm $SiO_2$/Si. AFM and Raman spectroscopy was used to confirm the layer number of the $MoS_2$ nanosheets. AFM image in Figure 1B shows that the $MoS_2$ flake consists of two thickness profiles, *i.e.*, 0.7 and 1.4 nm, as measured from its height profile in Figure 1G. It corresponds to the single- (1L) and double-layer (2L) $MoS_2$ nanosheets,[2, 3] which were further confirmed by their in-plane vibration ($E_{2g}^1$) and out-of-plane vibration ($A_{1g}$) modes in Raman spectroscopy (Figure 1F).[2, 3, 47]

MFM was used to characterize the obtained 1L and 2L $MoS_2$ nanosheets. Figure 1C shows the phase image of the same $MoS_2$ flake (Figure 1A) obtained simultaneously with the topography image (Figure 1B). Figure 1D and E show the MFM phase and amplitude images of the same flake at a lift height of 30 nm, respectively. Note that in order to avoid the response variation induced by different tips, all images were captured with the same tip. As shown in AFM phase image (Figure 1C), it is difficult to distinguish the 1L and 2L $MoS_2$ nanosheets since the difference of phase shift is very small (4.0º for 1L and 4.2º for 2L, Figure 1H). However, in the MFM phase image, obvious difference between the 1L and 2L $MoS_2$ nanosheets is observed (Figure 1D and I). 2L $MoS_2$ nanosheet has a bigger negative phase shift than does 1L $MoS_2$ (62 milli-degree (mº) for 2L and 47 mº for 1L), indicating that 2L $MoS_2$ has stronger attractive interaction with the MFM tip. Meanwhile, Figure 1E shows the MFM amplitude image, in which 1L and 2L $MoS_2$ nanosheets have positive amplitude shift. In the MFM measurement, the attractive force between tip and sample decreases the resonance frequency of the cantilever,[39, 40] resulting in the increase of vibration amplitude signal and decrease of phase signal (Figure S2 in SI). Therefore, the reverse contrast between MFM phase and amplitude images confirms that 1L and 2L MoS$_2$ nanosheets are magnetic. Furthermore, our MFM measurement of mechanically-exfoliated MoS$_2$ nanosheets at the decreased lift height gave a larger negative phase shift (Figure S3 in SI), which also confirmed that the mechanically-exfoliated MoS$_2$ nanosheets are magnetic. It is consistent with the previous reports on the MFM measurement of magnetic samples.[1, 42, 43, 45, 46]

As control experiments, the magnetic Fe$_3$O$_4$ nanoparticles (NPs) and nonmagnetic gold nanoparticles (Au NPs) were used to confirm the validity of the MFM measurement. As measured by the same MFM tip, magnetic Fe$_3$O$_4$ NPs also show negative phase shift in the combination with positive amplitude shift. However, nonmagnetic Au NPs show both positive phase and amplitude shifts (Figure S4 in SI), which is consistent with previous reports.[41, 42, 44, 45] In this case, the positive phase shift of Au NPs might come from the electrostatic interaction rather than the magnetic interaction.[42] Therefore, aforementioned MFM measurements are valid and able to serve as indication of magnetic response.

Importantly, our experimental results demonstrate that the magnetic response of MoS$_2$ nanosheets depends on their layer number. As shown in Figure 1D and I, the negative phase shift of 2L MoS$_2$ nanosheet increased ~32% (from 47 m$^\circ$ to 62 m$^\circ$) compared to 1L MoS$_2$. It further increased as the layer number of MoS$_2$ nanosheets increased to 7L. Figure 2A and C show the AFM topography and MFM phase images of a MoS$_2$ flake consisting of 1L and 7L MoS$_2$ nanosheets, respectively, which is confirmed by their thickness (1L: 0.8 nm, 7L: 4.7 nm,[48] Figure 2B). As shown in Figure 2D, the negative phase shift of 7L MoS$_2$ nanosheet increases by ~155% (from 22 m$^\circ$ to 56 m$^\circ$) compared to the 1L MoS$_2$ nanosheet, indicating the increase of negative phase shift as the layer number increased. Moreover, further increase of negative phase shift is observed as the thickness of MoS$_2$ nanosheet increased from 5.3 nm (8L) to 16 nm (~24L) (Figure 2E-H). There is no noticeable phase shift of MoS$_2$ nanosheets (from 283 m$^\circ$ to 271 m$^\circ$) when their thickness increases from 16 nm (~24L) to 43 nm (~66L) (Figure 2E-H). However, the negative phase shift of MoS$_2$ nanosheets decreases by 18% (from 283 m° to 231 m°) when their thickness increases from 16 nm (~24L) to 45 nm (~69L) (Figure 2E-H). If the thickness further increases to ~183 nm, the MoS$_2$ flake shows very weak positive phase shift and positive amplitude shift, implying the thick MoS$_2$ flake might be nonmagnetic or has no detectable magnetic response (Figure S5 in SI). This is consistent with a previous report in which it was found that the CVD-grown MoS$_2$ flakes with thickness of more than 100 nm showed much weaker magnetism than did the thinner MoS$_2$ flakes (thickness of ~20 nm).[27]

Moreover, MFM can be also used to characterize the single- and few-layer graphene nanosheets. As shown in Figure 3A, the mechanically-exfoliated 1L, 2L, 3L and 5L graphene nanosheets were successfully deposited on 90 nm SiO$_2$/Si, which were confirmed by Raman spectroscopy[49, 50] (Figure 3B) and AFM height measurement[51, 52] (see Figure 3C and the height profile in Figure 3D), respectively. The corresponding MFM phase image of graphene nanosheets (Figure 3E) shows that the 1L graphene has the strongest negative phase shift (~76 m°), which is larger than that of 2L graphene (~20 m°) and 3L graphene (~5 m°) (see the MFM phase shift profile in Figure 3F). However, the 5L graphene nanosheet exhibits almost no phase shift difference from the substrate (see Figure 3E and the MFM phase shift profile in Figure 3F). Furthermore, thicker graphene flake (*e.g.* 3 nm thick) showed similar result with the 5L graphene (Figure S6 in SI). All these results are consistent with a previous study, which reported that the thinner graphene has larger magnetic signal.[28] Note that graphene nanosheets showed different thickness-dependent magnetic response compared to MoS$_2$ nanosheets, *i.e.*, the negative phase shift of graphene nanosheets decreases as the layer number increases (Figure 3E and F).

It is well known that the magnetic contrast observed in MFM is highly dependent on the lift height.[39, 40, 42, 46] In order to fully characterize the magnetic response of 1L graphene, the MFM phase shift measurement was performed at various lift heights from 25 to 150 nm. Figure 4A shows the AFM image of 1L graphene and the corresponding MFM phase images with lift heights of 150, 100, 80, 50, 30 and 25 nm, respectively (Figure 4B-G). Obviously, the phase shift of 1L graphene is strongly dependent on the lift height. As shown in Figure 4H, the negative phase shift exponentially increases with decreasing the lift height, which is consistent with previous reports.[42, 46]

It is worth noting that the aforementioned results are based on the mechanically-exfoliated high-quality $MoS_2$ and graphene nanosheets. The solution-processed 2D nanosheets usually have different properties from those of mechanically-exfoliated ones. For example, the mechanically-exfoliated 1L $MoS_2$ nanosheet exhibits *n*-type behavior,[2, 3, 8] while the solution-processed one exhibits *p*-type behavior.[1, 22] In this work, MFM was also used to characterize the solution-processed single-layer $MoS_2$[1] and graphene oxide (GO) nanosheets.[53, 54] Figure 5 shows the AFM topography (Figure 5A and G), MFM amplitude (Figure 5B and H) and MFM phase (Figure 5C and I) images of solution-processed $MoS_2$ and GO nanosheets, respectively. AFM measurement indicates that the heights of $MoS_2$ and GO nanosheets are 1.3 and 1.4 nm (Figure 5D and J), respectively, confirming that they are single-layer nanosheets, which are consistent with previous reports.[1, 54-58] As shown in Figure 5F, the MFM phase measurement of single-layer $MoS_2$ nanosheet shows that $MoS_2$ nanosheet has the positive phase shift (~54 m°), indicating that solution-processed single-layer $MoS_2$ nanosheets might have the repulsive interaction with MFM tip, which is different from the result of mechanically-exfoliated single-layer $MoS_2$ nanosheets (Figure 1D and Figure 2C). Furthermore, the MFM amplitude measurement shows the weak negative amplitude shift (18 mV, Figure 5E), which confirms that solution-processed single-layer $MoS_2$ nanosheets have reverse magnetic signal to the mechanically-exfoliated single-layer $MoS_2$ nanosheets (Figure 1D-E). This difference might arise from the residual lithium on the solution-processed $MoS_2$ nanosheets, which also exhibited the *p*-type doping behavior different from the *n*-type doping behavior of mechanically-exfoliated ones.[1-3, 22] In addition, the GO nanosheets prepared by the solution method[53, 54] have not shown any detectable magnetic response (Figure 5H-I and K-L), which might be attributed to the presence of functional groups and defects.

**CONCLUSION**

In summary, for the first time, the magnetic force microscopy (MFM) is used to characterize the mechanically-exfoliated single- and few-layer $MoS_2$ and graphene nanosheets. By analysis of phase and amplitude shifts, the magnetic response was found in single- and few-layer $MoS_2$ and graphene nanosheets. Both $MoS_2$ and graphene nanosheets showed the thickness-dependent magnetic response. The magnetic response of $MoS_2$ nanosheets increased as the thickness increased from 0.8 to 16 nm, but decreased as the thickness further increased. However, too thick $MoS_2$ flake (> 183 nm) has no detectable magnetic response. In contrary to $MoS_2$ nanosheets, the negative phase shift of graphene nanosheets decreases as the layer number increased. The strongest negative phase shift of graphene nanosheets was found in the single-layer graphene. Even 5L graphene nanosheet showed almost no detectable phase shift. Compared to the mechanically-exfoliated $MoS_2$ and graphene nanosheets, the solution-processed single-layer $MoS_2$ nanosheets showed reverse magnetic signal while GO nanosheets exhibited no detectable magnetic response. Our MFM measurement of $MoS_2$ and graphene nanosheets opens up a useful means for the fundamental understanding of the intrinsic properties of 2D nanomaterials.

**METHODS AND MATERIALS**

***Preparation and MFM measurements of $MoS_2$ nanosheets, graphene nanosheets, $Fe_3O_4$ nanoparticles (NPs) and Au NPs.***

Natural graphite (NGS Naturgraphit GmbH, Germany) and $MoS_2$ crystals (SPI Supplies, USA) were used for preparation of mechanically-exfoliated single- and few-layer graphene and $MoS_2$ nanosheets, respectively, which were then deposited onto the freshly cleaned 90 nm

SiO$_2$ coated Si substrates (90 nm SiO$_2$/Si).[2, 3] The optical microscope (Eclipse LV100D, Nikon) was used to locate and image the single- and few-layer graphene and MoS$_2$ nanosheets.

Fe$_3$O$_4$ and Au NPs were prepared by using the previously reported method.[55, 59] Fe$_3$O$_4$ NPs is dispersed in hexane and spin-coated on a 90 nm SiO$_2$/Si substrate. Au NPs were deposited on a (3-aminopropyl)triethoxysilane-modified 90 nm SiO$_2$/Si substrate.[55]

Magnetic force microscopy (MFM) was carried out with a commercial AFM instrument (Dimension ICON with NanoScope V controller, Bruker) equipped with a scanner (90 × 90 µm$^2$) under ambient conditions. Si cantilevers coated with a cobalt/chromium film with the normal resonance frequency of 75 kHz and spring constant of 2.8 N/m (MESP, Bruker) were used for MFM images. The coating produced a coercivity of approximately 400 Oe. Other probes with similar properties (PPP-MFMR, Nanosensors®) were also tested and gave similar results. During our MFM measurements, the lift height is 30 nm if there is no specific clarification.

***Raman measurement of MoS$_2$ and graphene nanosheets.*** Analysis of the single- and few-layer MoS$_2$ and graphene nanosheets by Raman spectroscopy was carried out on a Renishaw inVia Raman microscope. All spectra were excited at room temperature with laser light (λ= 532 nm) and recorded through the 100× objective. A 2400-lines/mm grating provided a spectral resolution of ~1 cm$^{−1}$. The Raman spectra were calibrated by using the peak (520 cm$^{-1}$) of Si substrate.

*Conflict of Interest:* The authors declare no competing financial interest.

*Supporting Information Available.* MFM images of thick MoS$_2$ and graphene flakes, Fe$_3$O$_4$ NPs, and Au NPs. MFM phase, amplitude and frequency images of magnetic recording tape. This material is available free of charge *via* the Internet at http://pubs.acs.org.

*Acknowledgment.* This work was supported by MOE under AcRF Tier 2 (ARC 10/10, No. MOE2010-T2-1-060), Singapore National Research Foundation under CREATE programme: Nanomaterials for Energy and Water Management, and NTU under Start-Up Grant (M4080865.070.706022) in Singapore.

**Corresponding Author**

*E-mail: hzhang@ntu.edu.sg, hzhang166@yahoo.com.

Phone: +65-6790-5175. Fax: +65-6790-9081

Website: http://www.ntu.edu.sg/home/hzhang/

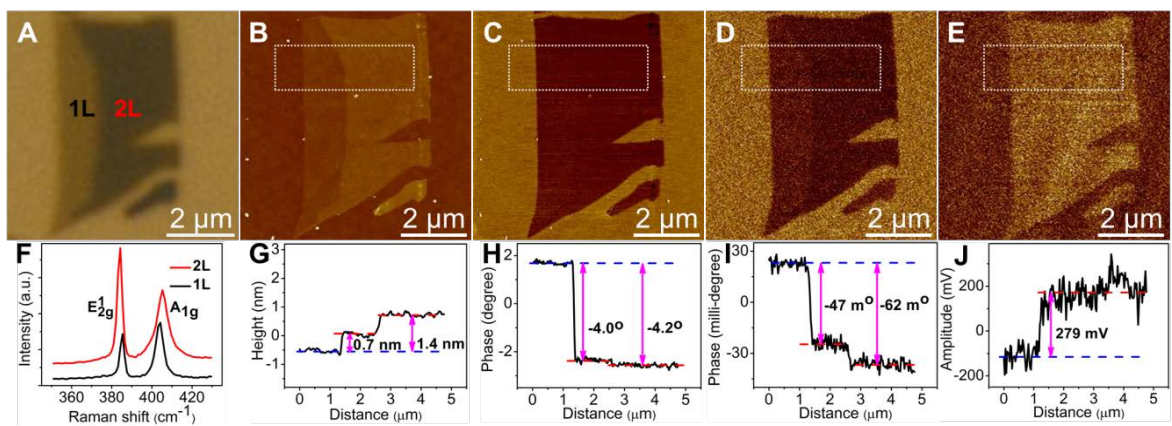

**Figure 1.** (A) Optical, (B) AFM topography, (C) phase, (D) MFM phase and (E) MFM amplitude images of 1L and 2L MoS$_2$ nanosheets on 90 nm SiO$_2$/Si. (F) Raman spectra of 1L and 2L MoS$_2$ nanosheets. (G-J) The corresponding profiles of the dashed reactangles in (B-E). The lift scan height is 30 nm.

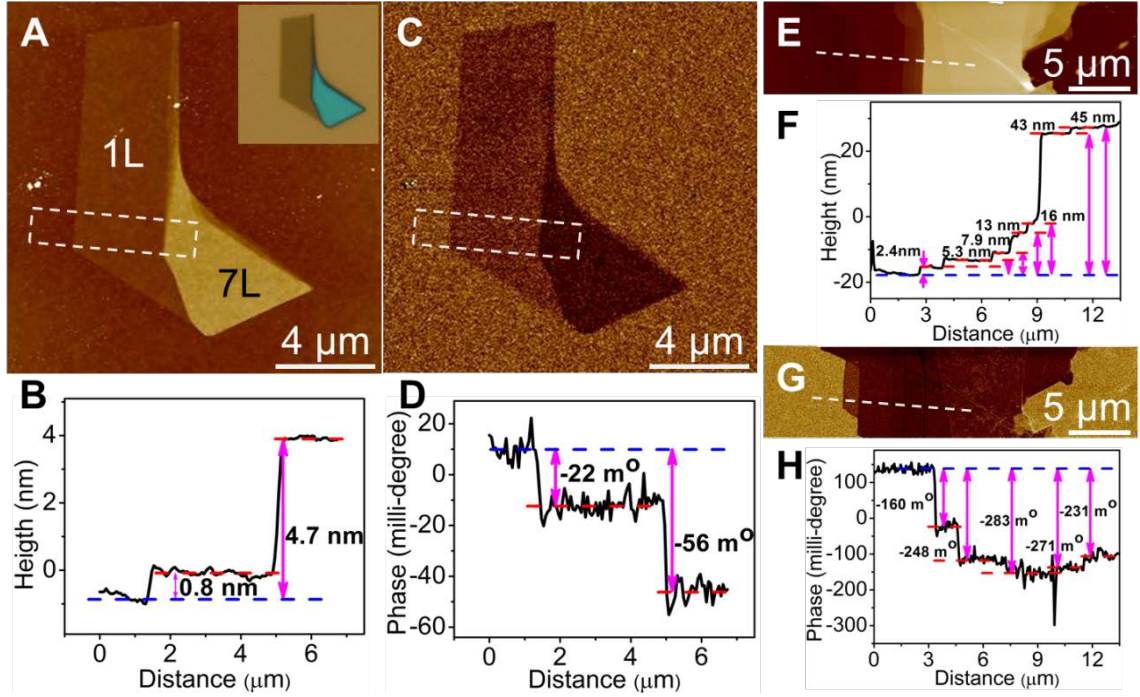

**Figure 2.** (A) AFM topography and (C) MFM phase images of 1L and 7L MoS$_2$ nanosheets on 90 nm SiO$_2$/Si. Inset in (A) is the optical image of MoS$_2$ flake. (B, D) The corresponding profiles of the dashed reactangles in (A) and (C), respectively. (E) AFM topography and (G) MFM phase images of thick MoS$_2$ flake. (F, H) The corresponding profiles of the dashed lines in (E) and (G), respectively. The lift scan heights for (C-D) and (G-H) are 30 and 20 nm, respectively.

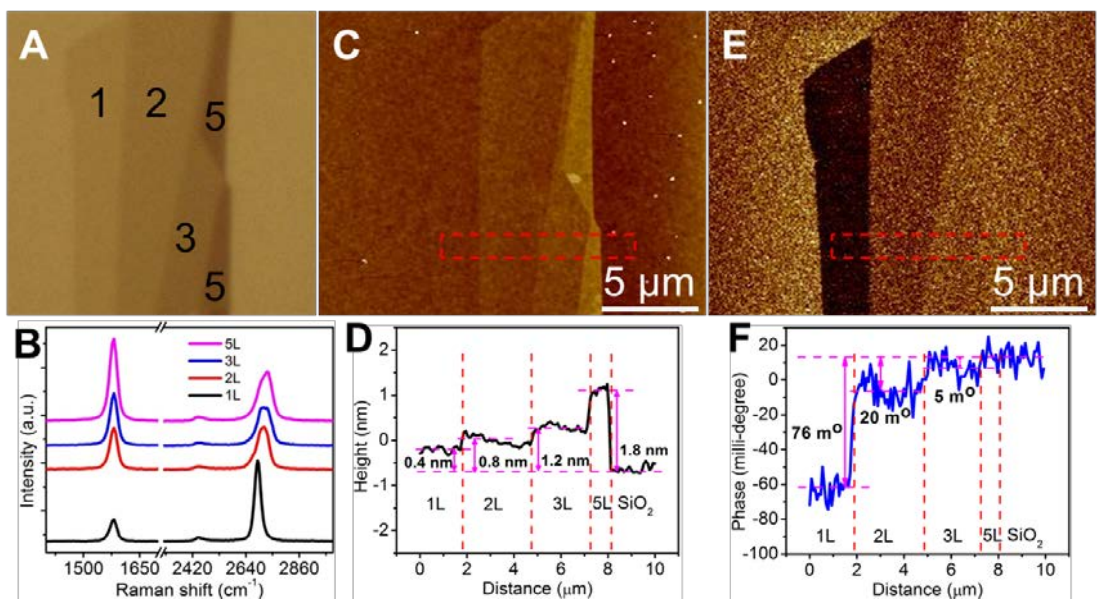

**Figure 3.** (A) Optical, (C) AFM and (E) MFM phase images of graphene nanosheets on 90 nm $SiO_2$/Si. Numbers in (A) indicate the layer number of graphene, which are confirmed by Raman spectroscopy and AFM height profile shown in (B) and (D), respectively. (D) AFM height profile and (F) MFM phase shift profile of the dashed rectangles in (C) and (E), respectively. The lift scan height is 30 nm.

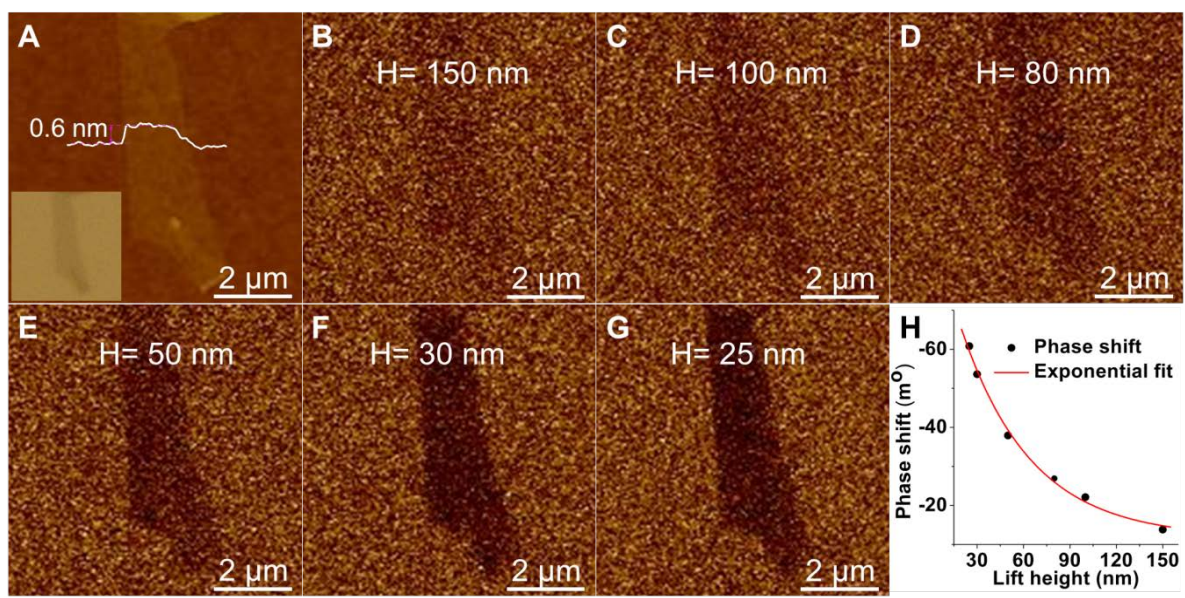

**Figure 4.** (A) AFM topography and (B-G) MFM phase images of 1L graphene on 90 nm SiO$_2$/Si at various lift heights: (B) 150, (C) 100, (D) 80, (E) 50, (F) 30 and (G) 25 nm. Inset in (A): Optical image of the 1L graphene on 90 nm SiO$_2$/Si. (H) The plot of phase shift *vs.* lift height obtained in the MFM measurement on 1L graphene. Red curve is the exponentially fitted curve.

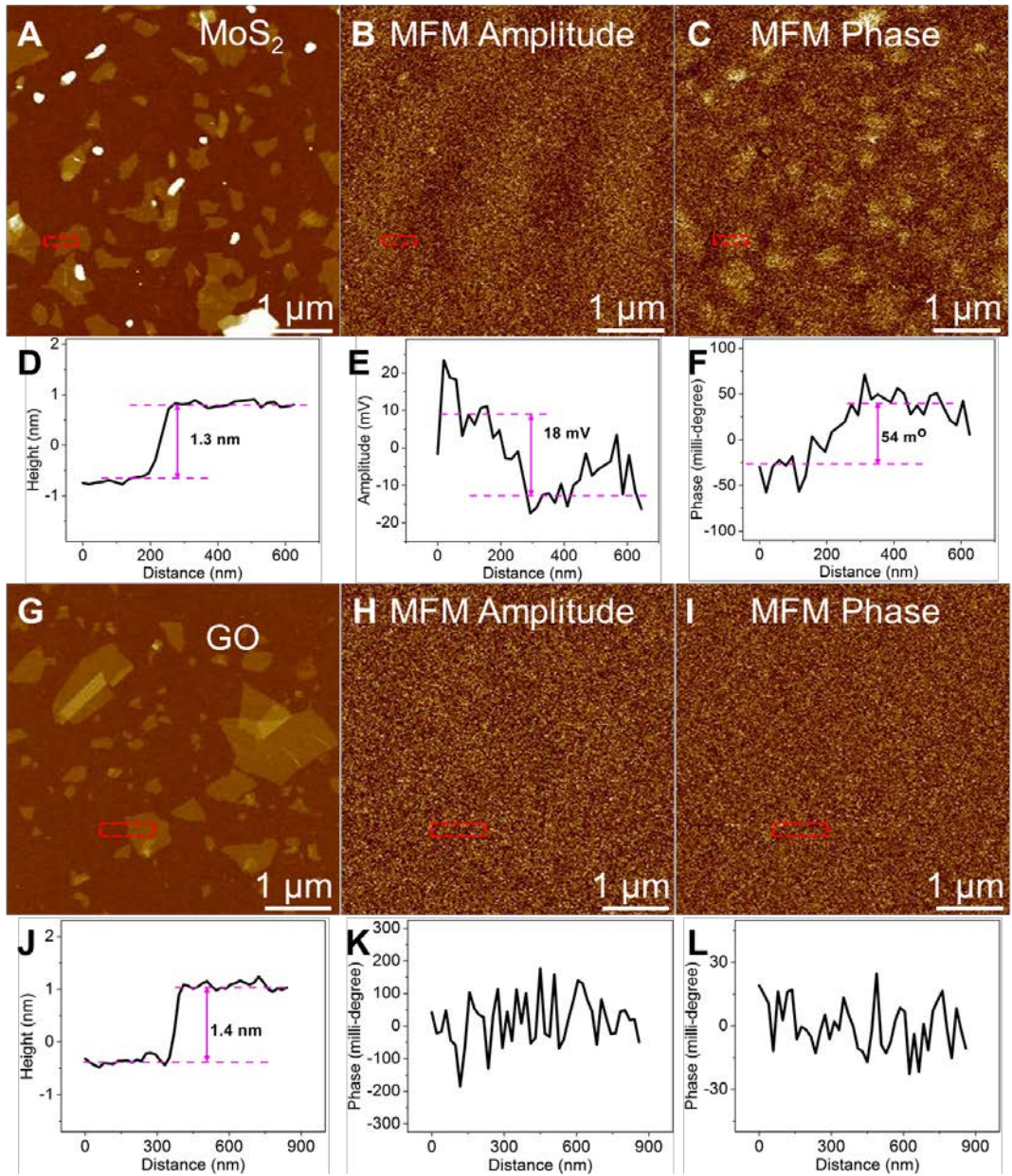

**Figure 5.** (A and G) AFM topography, (B and H) MFM amplitude and (C and I) MFM phase images of solution-processed single-layer MoS$_2$ (A-C) and GO (G-I) nanosheets, respectively. (D and J) AFM height, (E and K) MFM amplitude shift and (F and L) MFM phase shift profiles of the dashed red reactangles in (A-C) and (G-I), respectively. Note that the single-layer MoS$_2$ nanosheet shows negative amplitude shift and positive phase shift, while GO sheets have almost no MFM amplitude and phase shift difference from substrate. The lift scan height is 30 nm.

**ToC figure**

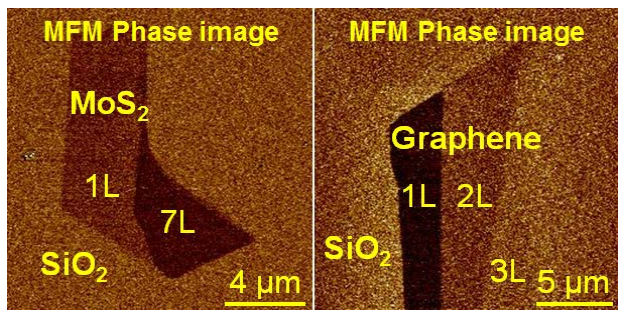

**Supporting Information**

# Investigation of MoS$_2$ and Graphene Nanosheets by Magnetic Force Microscopy

*Hai Li,[1] Xiaoying Qi,[2] Jumiati Wu,[1] Zhiyuan Zeng,[1] Jun Wei,[2] Hua Zhang[1]\**

[1]School of Materials Science and Engineering, Nanyang Technological University, 50 Nanyang Avenue, Singapore 639798, Singapore

[2]Singapore Institute of Manufacturing Technology, 71 Nanyang Drive, Singapore 638075, Singapore.

\* To whom correspondence should be addressed.

Phone: +65-6790-5175. Fax: +65-6790-9081

E-mail: hzhang@ntu.edu.sg, hzhang166@yahoo.com

Website: http://www.ntu.edu.sg/home/hzhang/

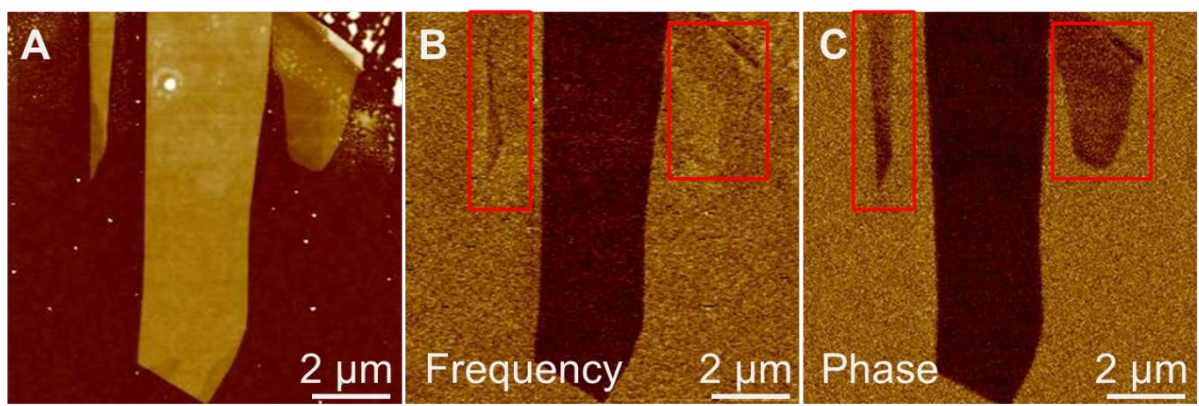

**Figure S1.** (A) AFM topography image of MoS$_2$ nanosheets. (B-C) The corresponding MFM images in frequency (B) and phase (C) channels, respectively. As shown by the red rectangles marked in (B) and (C), MFM phase signal is more sensitive compared to the frequency signal. The lift height is 30 nm.

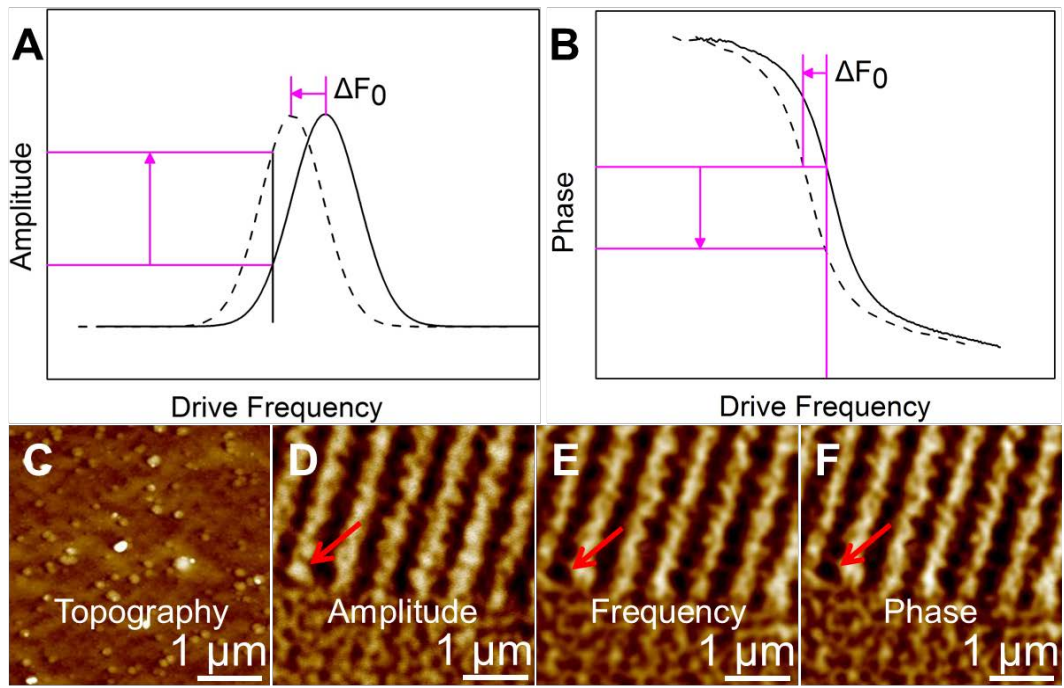

**Figure S2.** Schematical illustration of decreasing resonance frequency of cantilever induced by the attractive magnetic interaction between tip and sample, which is measured in (A) amplitude and (B) phase shifts at a fixed drive frequency. The decrease in resonance frequency results in the increase of vibration amplitude signal and decrease of phase signal. (C) AFM topography, (D) MFM amplitude, (E) frequency and (F) phase images of a typical magnetic recording tape. Note that the domain indicated by red arrow shows positive shift in amplitude image, but negative shifts in frequency and phse images.

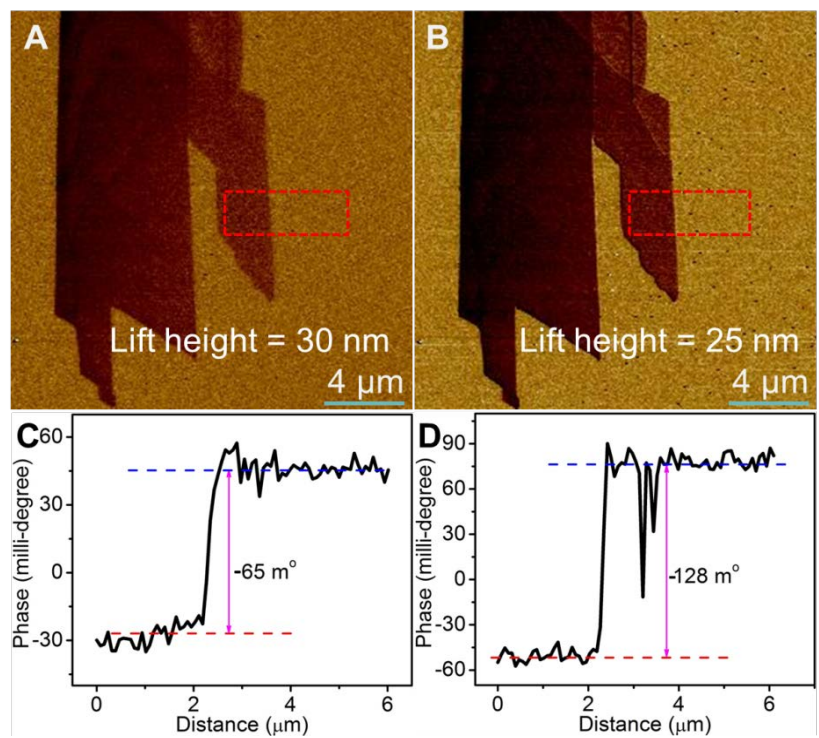

**Figure S3.** MFM images of MoS$_2$ nanosheets at different lift height: (A) 30 nm and (B) 25 nm. (C, D) The corresponding profiles of the dashed red rectangles in (A) and (B), respectively. Note that the stronger negative phase shift shows in the MFM image at smaller lift height, which further confirms the magnetism of atomically-thin MoS$_2$ nanosheets.

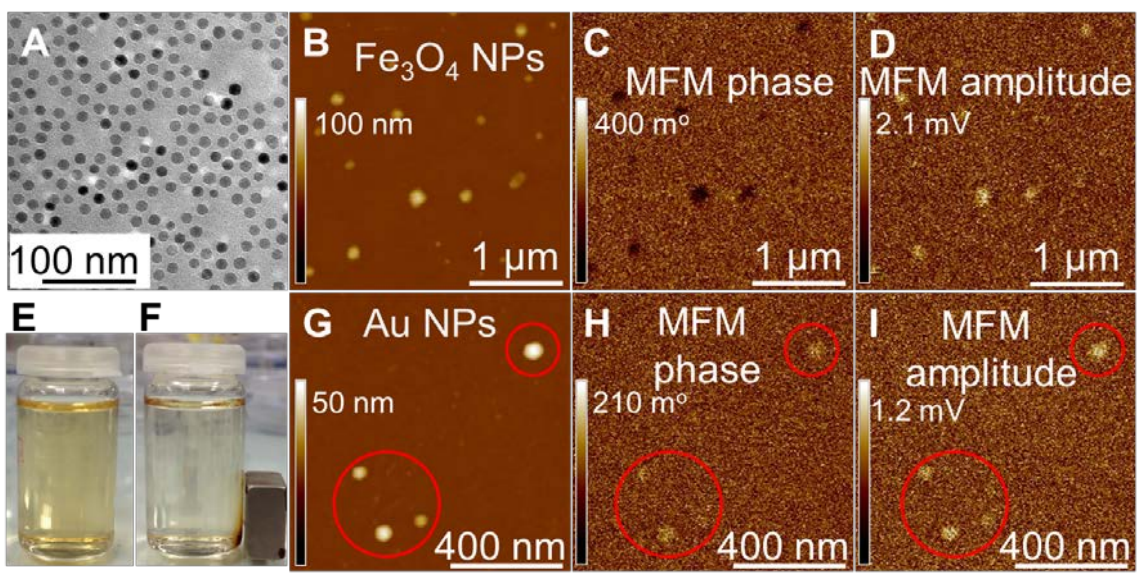

**Figure S4.** (A) TEM, (B) AFM topography, (C) MFM phase and (D) MFM amplitude images of $Fe_3O_4$ NPs. Note that the magnetic $Fe_3O_4$ NPs have negative phase shift and positive amplitude shift. (E) $Fe_3O_4$ NPs dispersed in hexane. (F) $Fe_3O_4$ NPs in (E) are attracted by a magnet and aggregate in the bottle bottom near the magnet, indicating they are magnetic. (G) AFM topography, (H) MFM phase and (I) MFM amplitude images of Au NPs. Note that the nonmagnetic Au NPs have both positive phase and amplitude shifts, which are marked by red circles. The lift height is 30 nm.

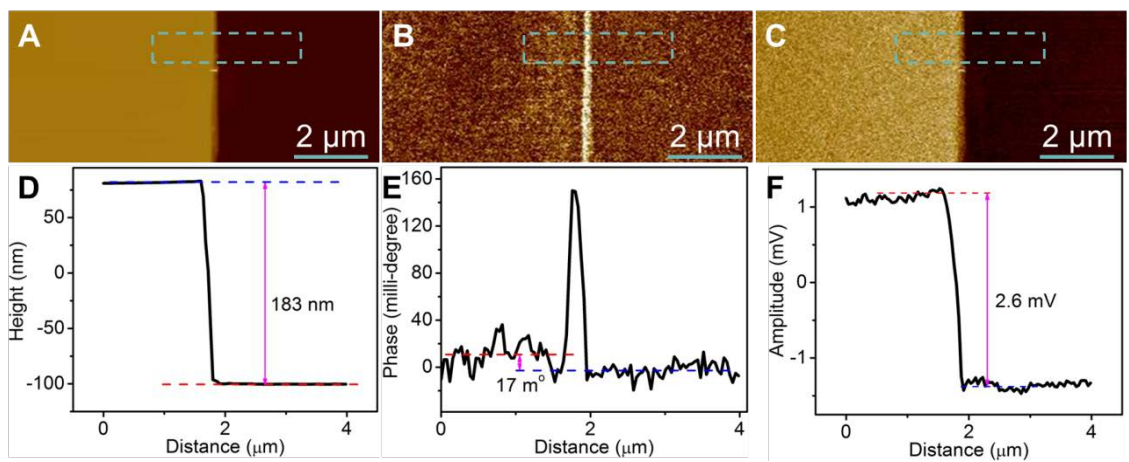

**Figure S5.** (A) AFM topography, (B) MFM phase, and (C) amplitude images of the MoS$_2$ flake with thickness of ~183 nm. (D-F) The corresponding profiles of the dashed rectangles in (A-C), respectively. The thick MoS$_2$ flake shows a weak positive phase shift (17 m°) and positive amplitude shift (2.6 mV) in MFM measurement, indicating it is nonmagnetic. The lift height is 30 nm.

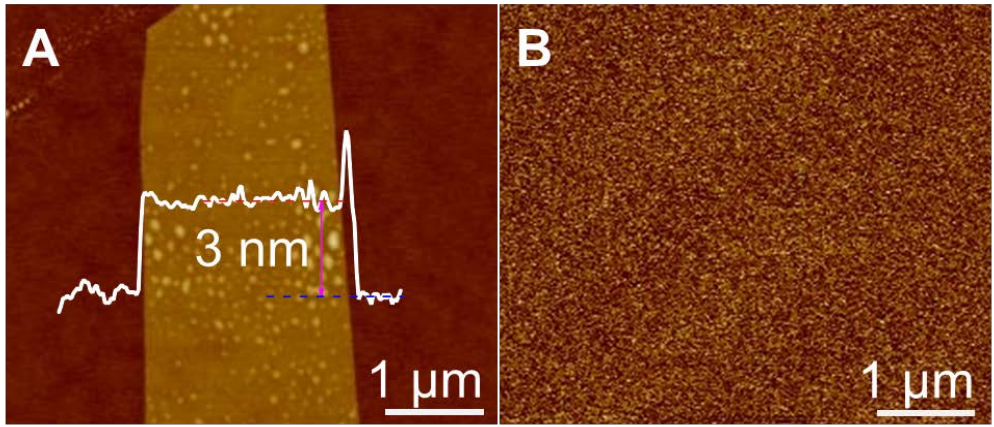

**Figure S6.** (A) AFM topography and (B) corresponding MFM phase images of graphene flake with thickness of ~3 nm. As shown in (B), this thick graphene flake shows almost no phase shift, indicating it is nonmagnetic or has no detectable magnetic signal. The lift height is 30 nm.